# AN INTERACTIVE ZOO GUIDE: A CASE STUDY OF COLLABORATIVE LEARNING

Hao Shi

School of Computer Engineering and Science, Victoria University,
Melbourne, Australia
hao.shi@vu.edu.au

## ABSTRACT

*Real Industry Projects and team work can have a great impact on student learning but providing these activities requires significant commitment from academics. It requires several years planning implementing to create a collaborative learning environment that mimics the real world ICT (Information and Communication Technology) industry workplace. In this project, staff from all the three faculties, namely the Faculty of Health, Engineering and Science, Faculty of Arts, Education and Human Development, and Faculty of Business and Law in higher education work together to establish a detailed project management plan and to develop the unit guidelines for participating students. The proposed project brings together students from business, multimedia and computer science degrees studying their three project-based units within each faculty to work on a relatively large IT project with our industry partner, Melbourne Zoo. This paper presents one multimedia software project accomplished by one of the multi-discipline student project teams. The project was called 'Interactive ZooOz Guide' and developed on a GPS-enabled PDA device in 2007. The developed program allows its users to navigate through the Zoo via an interactive map and provides multimedia information of animals on hotspots at the 'Big Cats' section of the Zoo so that it enriches user experience at the Zoo. A recent development in zoo applications is also reviewed. This paper is also intended to encourage academia to break boundaries to enhance students' learning beyond classroom.*

## 1. INTRODUCTION

The importance of a substantial software engineering project undertaken by undergraduate students is widely recognized and documented [1, 2]. Real Industry Projects and team work can have a great impact on student learning but providing these activities requires significant commitment from academics [3, 4]. In recent years, higher education institutions are under increasing pressure from government, industry and business organizations to equip students with required discipline based skills as well as generic skills. At the same time universities are experiencing cuts in resources and increased competition for student numbers [5]. Most computer science academics acknowledge the need for formal training in writing and speaking for students [6, 7]. These skills are also highly desired by computer scientists in industry and by employers. Even computer science students rate the need for oral, written, and interpersonal skills a close second to technical skills [6, 8]. Yet many universities do not require communications training at all as part of the curriculum, or what is offered does not go far enough [9].

The International journal of Multimedia & Its Applications ( IJMA ), Vol.2, No.2, May 2010

Victoria University is a multi-sector institution consisting of Higher Education and TAFE (Technical And Further Education) located on multiple campuses in the western region of Melbourne, Australia [10]. Currently there are three faculties in higher education sector, namely Faculty of Arts, Education and Human Development, Faculty of Business and Law, and Faculty of Health, Engineering and Science. About 40% students of Victoria University are from homes where English is not the first language and 50% do not have a family history of post-secondary education [11]. When developing a teaching program, academics must consider the specific needs of the student profile and prepare courses that will enhance learning outcomes and the range of graduate destinations. This program has been initiated and planned for several years in order to mimic the real world IT industry workplace in the University environment. The objective is to have business, multimedia and computer science students from three higher education faculties studying together in a joint project team to accomplish a relatively large IT project. The challenge is to break the boundaries between the three faculties to create collaborative teaching and learning environment beyond traditional settings with one teaching subject. Students work in cross disciplinary teams, supervised by staff from all three faculties that bring unit expertise from each to create robust outcomes that address the major needs identified by our partner organization, Melbourne Zoo. The students are drawn from three pre-existing units within each faculty that currently allow for group project work. The paper, presents the accomplished multimedia software project, in terms of the project background, system overview, graphical user interface (GUI) design and the two GPS-based components, completed by one of the multi-discipline project teams, called 'Interactive ZooOz Guide'.

## 2. RECENT DEVELOPMENT

Although many wireless technologies become available during the last decade, there is very limited adoption to create an innovative zoo experience for their patrons. In 2003, the Aalborg Zoo had an attempt to adopt the BlueTags which allowed parents keep track of children at the zoo using an innovative Bluetooth tracking system [12]. At the entrance, parents rented the BlueTags' BodyTag for their children and register the required information along with contact details. They received an SMS on their mobile phone with a code to be used when making inquiries about the location of the individual child. The parent simply sent an SMS with the specific SMS code and, through the tracking software and the SMS gateway, a response came back with information about the child's location. Unfortunately the innovation doesn't become popular as it was used for providing security rather than main focus of Zoos' activities.

The GPS-enabled 'Interactive ZooOz Guide' documented in this paper was developed in 2007. Since then, many GPS-enabled projects were carried out. In 2008 the GPS Ranger was specially tailored to the patrons of the Zoo markets as Zoo Ranger. A Zoo Ranger is a handheld video tour guide system that provides informative and entertaining videos, audio, photography and animation about the Zoo's animals and exhibits based on location using the power of GPS technology. The zoo patrons can learn more about the Zoo's animal collection, enrichment activities and behaviour training from keepers and staff while on their tour. San Francisco Zoo was the first zoo in America to deliver Zoo Ranger tours as shown in Figure 1 [13]. The Oklahoma City Zoo was the third zoo in USA to offer this fun and unique touring device [14]. The zoo patrons can rent a Zoo Ranger tour at the Zoo's Strollers window.

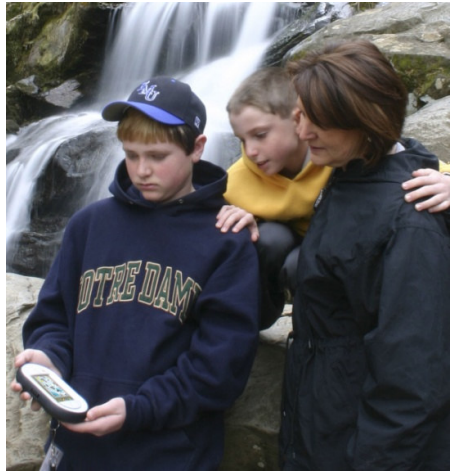

(a) Zoo Ranger at San Francisco Zoo [15]

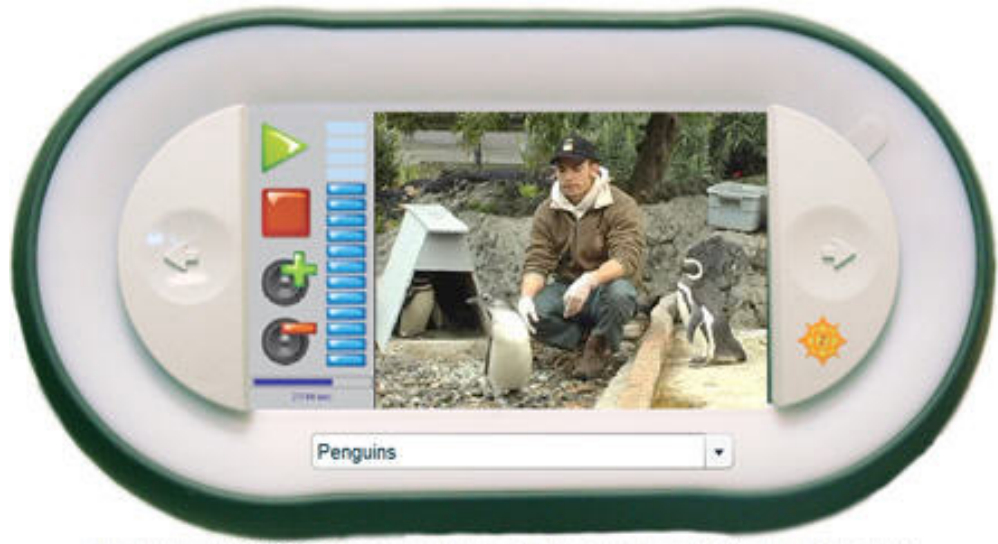

(b) A Handheld Zoo Ranger [16]

Figure. 1 GPS Ranger and Zoo Ranger

The real breakthrough was made in 2009 when iphone applications were developed for Zoos. Woodland Park Zoo released the first iPhone application [17] that allows visitors to track own location on zoo grounds, discover more about the animals, and access daily activity schedules to make the most of the next zoo visit as shown in Figure 2.

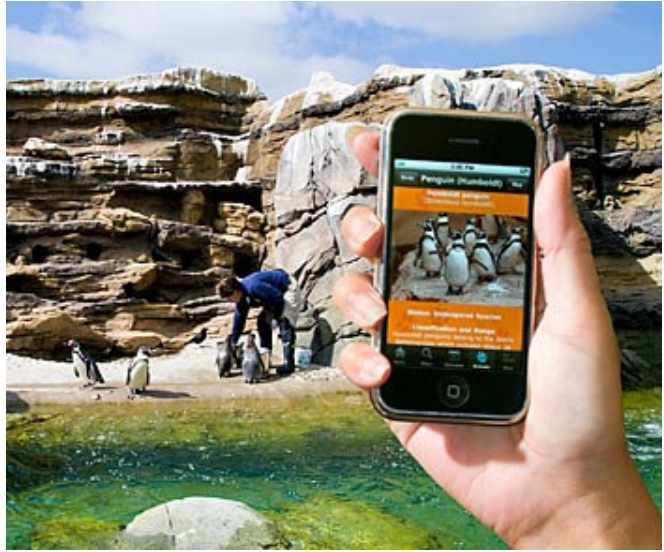

Figure 2. The Woodland Park Zoo iPhone application [18]

The Houston Zoo is the 2nd zoo in the world after the Woodland Park Zoo in Seattle, to offer such a service [19]. The Houston Zoo's new free iPhone application displays patrons' location on Zoo grounds using real-time GPS coordinates and allows visitors to access photos and videos of Zoo exhibits and animals and access daily 'Meet the Keeper Talks' and presentations.

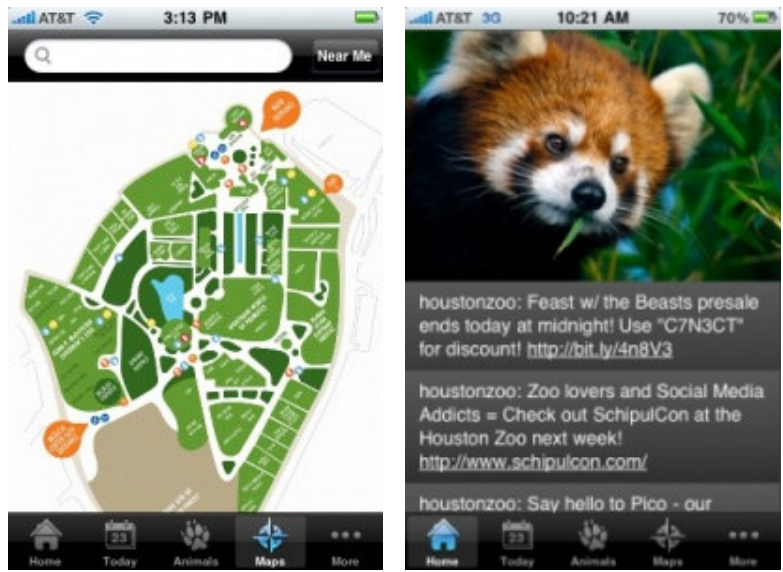

Figure 3 Screen Snapshots of iphone applications [19]

Although iphone application provided very promising future for Zoo applications but the practical challenges still remains, e.g. WiFi connection and coverage.

## 3. BACKGROUND

Our industry partner, Melbourne Zoo has more than 320 wildlife animals on display everyday during the year in a wonderful botanical setting only 4km from the Melbourne CBD, Australia [20]. There are many special attractions at the Zoo such as the African Rainforest, Asian Rainforest, Great Flight Aviary, the Big Cats Close-up, the Fur Seals Pool and Penguins Pool. The Zoo has considered the use of the latest technologies but it lacks resources and funding to implement their ideas. The concept of incorporating interactive technology as part of the Zoo experience has become the driving force behind their decision to sponsor the Interactive ZooOz Guide project so that it can make the Zoo experience much more enjoyable for their patrons.

Currently each patron is given a paper map as shown in Figure 4 [21] where they have to figure out which direction they are facing to navigate through the Zoo. The major complaint from the patrons is that the map is confusing to read, they get lost quite easily within the Zoo and in the end they don't get to see all the animals. In order to improve the navigation, the Zoo has created an Interactive Flash Map [22] as shown in Figure 5 which is available online at its website. Unfortunately the interactive maps is good for planning purpose but it doesn't provide a practical solution to the patrons once on ground inside the Zoo. As a result, the Interactive ZooOz Guide is proposed with the main goal to demonstrate the possibility of running interactive multimedia contents on a personal GPS (Global Positioning System)-enabled PDA (Personal Digital Assistant) device.

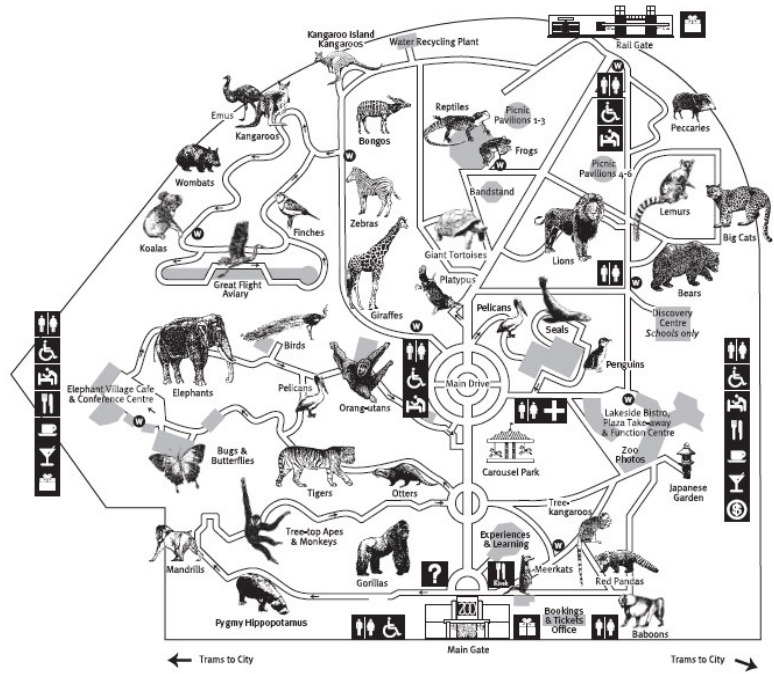

Figure 4. Melbourne Zoo Map [21]

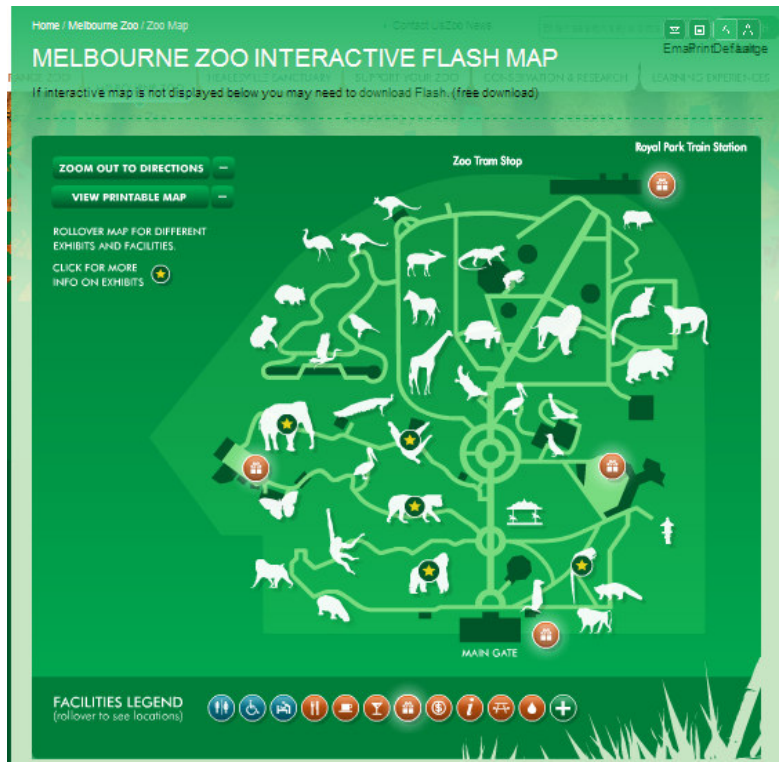

Figure 5. Melbourne Zoo Interactive Flash Map [22]

## 4. SYSTEM OVERVIEW

The Interactive ZooOz Guide is completed by three groups of students:

- feasibility study and economic analysis conducted by business students,
- multimedia contents generated by multimedia students, and
- system design and implementation developed by computer sicence students

Since it is at a research stage, the Interactive ZooOz Guide is focused on the navigation of the Zoo and multimedia contents delivery at the hotspots of the 'Big Cats' section of the Zoo where tigers, lions, jaguars, and leopards are living. The Zoo would like the ZooOz Guide to target the general public but specifically tailored for teachers and students. It aims to make it easier for their patrons to navigate around the Zoo. Also, it needs to be educational, which they can receive detailed information about animals when they reach the hotspots.

The system consists of three key components:
- PDA device supporting the main program
- A GPS receiver
- A relational database

The GPS receiver is connected to the PDA via Bluetooth. The system allows its users to check their connectivity with the GPS receiver and then provides the current location, i.e. longitude and latitude of the PDA device.

## 5. GRAPHICAL USER INTERFACE DESIGN

### 5.1. Startup Screen

The startup screen is designed by the computer students. It incorporates the Melbourne Zoo's logo and features a leopard at the background to represent the 'Big Cats' as shown in Figure 6. The startup screen stays for 5 seconds to allow the system to get ready. Once the startup screen disappears, the program identifies the COM port number of the GPS Bluetooth device, establishes the communications with the GPS satellites via the GPS receiver as shown in Figure 7, and displays the main screen. If the program fails to communicate with the GPS satellites, the users can "restart" GPS connection. If still unsuccessful, the program automatically exits.

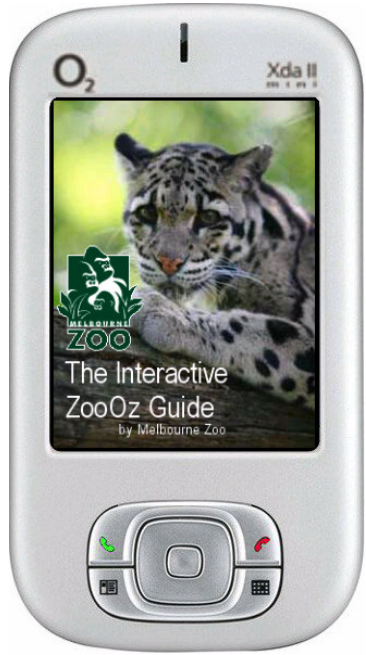 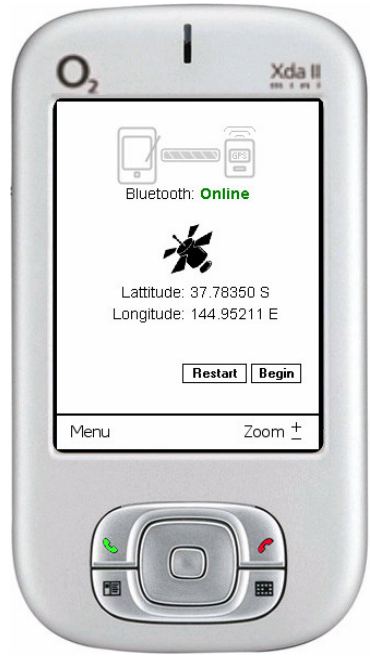

Figure 6. Startup Screen [23]      Figure 7. Connection Screen [23]

### 5.2 Menu Screen and Submenu Screens

There are two functionalities available: Menu and Zoom, which can be seen from the main screen in Figure 8. Once the menu is clicked, it shows the six submenus as illustrated in Figure 9. Then the user can select one of the six options from Check Connection, Show Coordinates, Tour Guide, Search, Events (timetable) and Close.

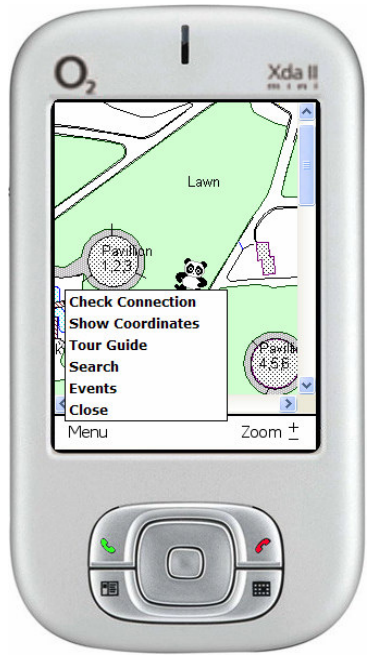 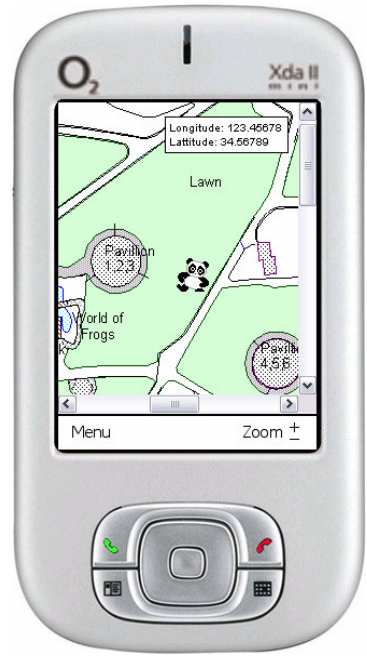

   Figure 8. Menu Screen [23]   Figure 9. Coordinates Screen [23]

## 6. THE INTERACTIVE ZOO GUIDE

There are two major GPS-based software components in the project. The first part is an interactive map which allows the patrons to navigate through the Zoo while the second part is the multimedia contents for the hotspots at the 'Big Cats' section of the Zoo.

The interactive map uses a high resolution of CAD (Computer Aided Design) picture provided by Melbourne Zoo as a background layer as shown in Figure 10. Layered on top of the map are hotspots which are defined by a set of coordinates. Then the 3rd layer is the cursor image which remains uniform, unaffected by the other two layers.

The range of readable longitude and latitude coordinates are unlimited, although the interactive map only shows if the user of the PDA is in the range of the Zoo. The map image shifts left or right and moves up or down in real-time according to the user's request, for both direction and distance. It enables the user to navigate around the Zoo and guide himself/herself easily with the help of a 'Panda' cursor displayed in the centre of the map. The "+" green *crosses* represent the points of interest within the range, where the user can trigger an action to retrieve the hotspot information. In addition the Zoom in/out function allows the users to set the desired size by pressing '+' and '-'.

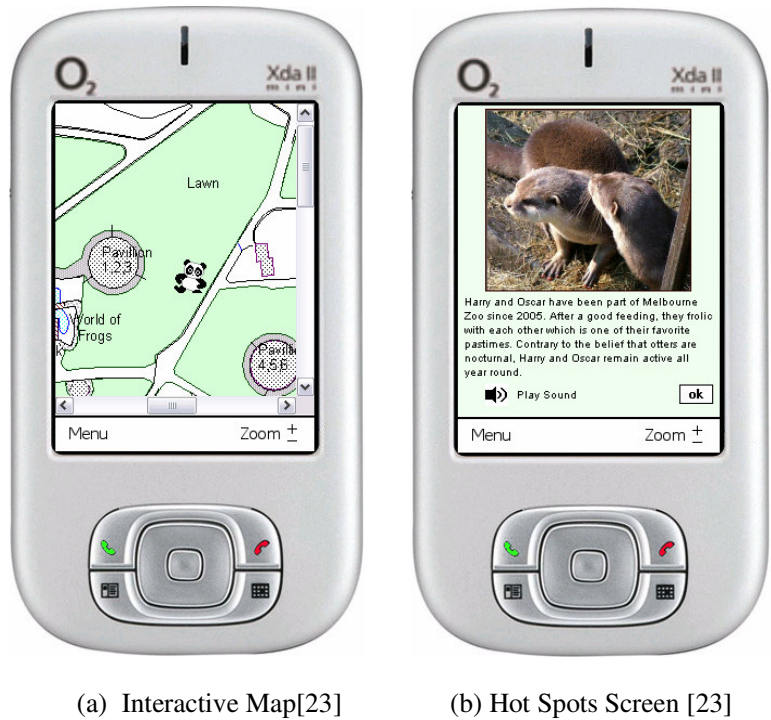

(a)  Interactive Map[23]        (b) Hot Spots Screen [23]

Figure 10. Interactive Zoo Guide

## 7. CONCLUSIONS

The proposed project brings together staff and students from the three higher education faculties of the University. The students from business, multimedia and computer science degree courses team up and form one multi-discipline team working on the Interactive ZooOz Guide project sponsored by the Melbourne Zoo. The team produces two major GPS-based components: the interactive Map and the multimedia information of animals on the hotspots at the 'Big Cats' section of the Zoo. The field tests have proved that the interactive map makes navigation around the Zoo much easier than the paper-based map and multimedia contents and hotspots provide unique rich experience for the patrons at the Zoo. Although the project is still a research prototype but it has set a solid foundation for the future development. It is very pleasing to see the students working together and successfully completing a relatively large project, which cannot be achieved by a single discipline team. The project has also created a collaborative learning and teaching environment, which have great impact on students' learning.

## ACKNOWLEDGEMENTS

The author would like to thank our industry partner, Rick Hammond from the Melbourne Zoo, colleagues, Megan Chudleigh, Associate Professor Jordan Shan, Luke Low, Dr. Donna Dwyer for their collaboration in teaching. Special thanks go to our computer science project students, Thang Nguyen (team leader) and Steven Manceski for their major contributions in design and development of the ZooOz Guide, the multimedia students, Stephanie Wulf, Vanessa Jalovec and Jaesel Magallanes for some multimedia contents and the business students, George Fotinos and Travis Mollica for their initial involvement. This project was proudly sponsored by Melbourne Zoo, Australia and funded by Teaching and Learning Support (TLS) grants from Victoria University, Melbourne, Australia.

**Author**

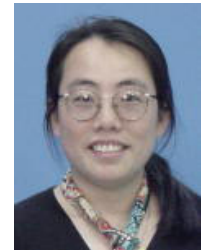

Dr. Hao Shi is an Associate Professor and the ICT Project Coordinator at School of Engineering and Science, Victoria University, Australia. She completed her PhD in the area of Computer Engineering at the University of Wollongong and obtained her Bachelor of Engineering degree from Shanghai Jiao Tong University, China. She has been actively engaged in R&D and external consultancy activities. Her research interests include p2p Networks, Location-Based Services, Web Services, Computer/Robotics Vision, Visual Communications, Internet and Multimedia Technologies.